\documentclass[12pt]{article}
\usepackage{a4}
\usepackage{graphicx}
\usepackage{amsfonts}
\usepackage{amscd}
\usepackage{latexsym}
\usepackage{epsf}
\usepackage{amsmath}
\numberwithin{equation}{section}

\newcommand{\be}{\begin{equation}}
\newcommand{\ee}{\end{equation}}

\newcommand{\bea}{\begin{eqnarray}}
\newcommand{\eea}{\end{eqnarray}}

\newcommand{\f}{function }
\newcommand{\fs}{functions }
\newcommand{\eqq}{equation }

\newcommand{\eqqs}{equations }

\newcommand{\nn}{\nonumber}

\newcommand{\fr}{\frac}

\newcommand{\pd}{\partial}

\newcommand{\la}{ \lambda }

\newcommand{\om}{ \omega }

\newcommand{\vp}{\varphi }

\begin{document}

\title{On a Local Concept of Wave Velocities} 
\author{
{\sc I.~V.~Drozdov }\thanks{e-mail: drosdow@uni-koblenz.de} \\
{\small and}\\
A.~A.~Stahlhofen \thanks{e-mail: alfons@uni-koblenz.de}\\
\small  University of Koblenz, Institute for Natural Sciences\\ 
\small  Department of Physics\\
\small  Universit\"atsstr.1, D-56080 Koblenz, Germany}
\maketitle

\begin{abstract}
The classical far field concept of wave velocities has its merits while exibiting
 intrinsic difficulties.

 A general local approach for the definition of velocities
 and especially phase velocities for waves avoiding these difficulties is proposed. 
  It includes the classical definitions as particular cases
 and can be applied to waves
 of an arbitrary structure, and to arbitrary propagation media as well. Applications of the
  formalism  are elucidated and some basic properties
 of the local concept defined here are discussed.       
\end{abstract}

\section{ Some remarks on wave velocities}
\label{ preface}

 Waves are conventionally described via propagating harmonic \fs containing a periodic factor
  of the form $e^{i\vp}$.
   The argument of these periodic \fs, interpreted as the "phase", is usually 
    identified with $(t-x/c)$ via the free wave equation, where  $c$ is the 
    "phase velocity" \cite{brillouin}.
  
   This canonical approach originates from classical wave optics and analyzes ordinary 
   light waves propagating in the "far field", where the wave is far away from the source.
    Such a light wave is a solution
   of a source free wave \eqq described by the periodic \f mentioned above.
    The Ansatz earned his own merits in the context of several wave phenomena and in particular
    classical wave optics \cite{born}.  
  
   A second fundamental concept of characterizing wave propagation 
   is that of the "group velocity".
     This concept is - mathematically speaking - 
     a comprehensive definition for a specified (linear) superposition of solutions
     of the free wave equation with the same periodicity properties      
    usually expressed by the frequency distribution of the constituent   
    periodic waves.
     This definition 
     is also an inherent far field concept considering "source-free" waves propagating 
     in a strongly homogeneous and isotropic medium. This medium is characterised
      only by its "dispersion" (supposed to satisfy additionally the Kramers-Kronig relations), 
i.e. by a dependence of the complex index of refraction $n(\om)$ on the periodicity
 parameter of the wave - the frequency $\om$.

This basis for definitions of  "phase" and "group" velocities turns out to be appropriate
  only for the special class of wave propagation mentioned above.
   The notions of a signal and its velocity developed in this
 context did accomodate many experimental data (see e.g.\cite{shiren}).    
 It is however by no means obvious, to what extent these far field concepts can be applied to near field problems.
  This leads to the question, if this classical approach is flexible enough to cover more general wave phenomena
  apart from the "special cases" mentioned above. A possible answer to this question is the focus of the present
   work.\\ 
Before proceeding we list some actual problems accompanying the classical definitions and 
suggesting a resolution by a local definition of a propagation velocity \cite{sonnenschein} as proposed here.
 
Theories developed for general not necessary periodic electromagnetic pulses \cite{sherman, bloch},  
 did draw basically on the adopted "canonical" scenario.
  The analysis for a wave of an arbitrary form, for instance, 
  has been based in general (by analogy to the classical case)
  on a representation via periodic \fs 
  by means of Fourier analysis.  
   A well defined "phase velocity" is assigned to each Fourier component equipped with its frequency;
 a corresponding "group velocity"  has subsequently been defined for the complete Fourier superposition 
 (i.e. for the "wave packet"). 
 As a consequence, the classical wave theory (in all its facets) is based on apparently
 "canonical" definitions relying on the phase of periodical functions and of groups of such functions.
 Such a strategy treates this as the inherent kernel of wave phenomena.

 A close inspection, however, reveals mathematical deficiencies and several shortcomings with respect to
 physics thereby rendering the classical "canonical" approach to some extent non-natural as elucidated now. 
 
First of all, a propagating wave
   is per definition a local space-time distribution which is a solution of a local differential equation.
     Hence the definition of velocity as a space-time relation should respect this frame and
     can be expected to be local as well.

The canonical definitions of a velocity ignore this fact and involve
 a frequency and a wave
     vector, which are nonlocal parameters in following the space- and time-periodicity, 
assumed separately in each case.     
      Moreover, these parameters have nothing to do with the propagation process itself, since they do not enter in the wave
      equation. Basically, the wave equation admits a propagation of arbitrary shapes without presumption of
      periodicity. In other words, the canonical approach bears anyway on a representation of an arbitrary solution
      by a set of periodic harmonic functions, each of that does not obey, generally speaking, the original wave equation
    (i.e. is not a solution).   
     An attempt to apply it on waves resulting from non-linear \eqqs such as solitons for instance, demonstrates
     this problem in a clear way since in this case even a linear superposition of solutions is not anymore a solution! 
         
Second,   
  the subject of transmission and velocity of a signal is based on sufficiently local procedures of measurement \cite{guertler}.
  Roughly expressed, the intervals between several space-time points are measured, where the attribute under study is detected.  
 Definitions based on periodicity parameters can possess generically neither time- nor space-locality.
 Moreover, any Fourier transformation is basically a global object, and all manipulations concerned are
 in general mathematically exact only with integration over the whole space and whole time, as well as
 over an infinite frequency band for backward transformations.
 
   The "canonical" definitions of the "phase" and "group" velocity mentioned above \cite{brillouin}, are based on 
   the special case of a propagation of periodic harmonic waves in homogeneous media. Thus there is already an essential
    contradiction between the local character of wave propagation and the inherent globality contained in the definitions of 
    wave velocities.    
  
Any attempt to construct a local measurable object using Fourier sums or integrals contain an essential contradiction
as outlined above and provides indeed no real locality.
  This is the source of several problems arising when replacing originally local features by 
 "microglobal" ones \cite{sherman, kohmoto2002, bloch}.
 
  The validity of this approach has to be checked from the mathematical point of view in the sense
  of functional analysis as well as from physical consistence and 
     in fact it turns out to be correct only in special cases as mentioned above.  
     
 Even if the approach is supported by mathematical consistency 
 (like the assumption of an infinite frequency band \cite{brillouin}),
  it does not lend itself easily to a transparent physical interpretation an the calculations are cumbersome.
  
 These drawbacks, are for instance, the origin of all problems involved 
 in the theory of signal transmission in media 
 and an interpretation of the results.
   
 The classical papers since \cite{brillouin} and later improvements thereof \cite{bloch, oughstun},
  still contain an essential mismatch between local and global wave features, based on several misleading definitions
   (like signal velocity, energy velocity and group velocity \cite{bers, pacher}).
  This ansatz is bound to lead in applications to controversial results. For example, it seems not more to be surprising, 
  that the subject  of "group velocity" failes to describe a propagaton of ultra-short pulses \cite{kohmoto}.\\
      
  These remarks shall suffice to motivate some alternative concept of wave
   propagation and wave velocities for the following reasons:    
     
\begin{enumerate}          
\item The canonical "harmonic" approach represents local wave features in terms of generically non-local attributes
\cite{sonnenschein, schriemer}. Hence:
\item The mathematical equipment is not exact for finite physical values.
\item It is therefore difficult to impossible to apply it to for near-field effects, if the size of space-time regions
 are of an order of magnitude smaller than the wave periodicity parameter;
\item It is poorly suited for inhomogeneous and anisotropic media, as well as for ultra-short pulses
of arbitrary form and for nonlinear waves \cite{kohmoto2002, gomez}.  
\item As a consequence, the applicabilty of this approach for several fields of modern quantum optics, nanooptics
 and photonics is barely justified, since these topics deal with the parameter areas outside its range of 
 validity \cite{centini}.  
\end{enumerate} 
 
 Moreover there is a clear reason to avoid this "harmonic" approach from the pure technical
 point of view:    
 on the one hand it turns out to be an essential restriction of generality and universality of the
 theory; on the other hand it brings to play ( especially the representation via Fourier series and integrals )a number of
 problematical artifacts such as an apparent violation of causality, infinite frequency bands for signals
  and so forth.
 These problems enforce further theoretical constructs like the analysis of dispersion relations \cite{nussenzveig}
 to resolve these artifical problems.\\
  
 A first step towards a more general concept should be based on an independent alternative
  approach without doubting the canonical "harmonic" criteria, leading to the same (verified)
  results in known special cases.
 
 The aim of the present paper is to establish a transparent criterion for evaluation
  of the propagation velocity for a quite arbitrary signal, that does not need an explicit
   representation in terms of harmonic or exponential functions at all, and with minimal loss of
   generality in other respects.  
 First of all, we have to recall, what is being measured in experiments and what is meant
 when speaking about a "wave velocity", thereby providing the spectrum of wave velocities.
 This discussion is presented in Sec.2. 
 Sec.3 is devoted to a test of the definition using some simple well known examples.
  Some remarkable properties of the behaviour of the wave velocities defined here under 
  relativistic transformation are outlined in Sec.4. 
 The discussion is concluded in Sec.5 by a practical interpretation and conversion from a local to a global
  evaluation of signal velocities.

\section{ N-th order phase velocity (PV)}
\label{ PVs}

  The following discussion is restricted to a 1+1 dimensional space-time for simplicity. 
    To evaluate the speed of a moving matter point one has to check the change of the 
    space coordinate $\Delta x$ during the time interval $\Delta t$. How can this ansatz be applied
    to waves ?  
       
  An arbitrary wave is a \f defined on the 2-dimensional manifold $(x,t)$ and 
  one has no {\it a priori } defined fixed points to follow up as in the former case.
  Let us therefore analyze a continuous smooth \f $\psi(x,t)$ of arbitrary shape.\\
   
   In a first step towards a local definition of velocity we consider a vicinity ${\cal U}(x)$ of a 
    certain point $x$ at the certain time $t$. Further we assume the local information 
    about the \f $\psi(x,t)$ to be measurable, i.e. we should be able (at least in principle) to evaluate
    the values of the \f $\psi(x,t)$ itself and all its derivatives in the point $x$ as well,
 and in the vicinity  ${\cal U}(x)$.
 
  A description of propagation is based on
  monitoring the fate of a certain attribute ("labelled point") of this shape, i.e.
   one finds the same attribute at the next moment $t_1$ 
  in the next point $x_1$. 
 
 \begin{figure}[h]   
\epsfxsize=10cm\epsffile{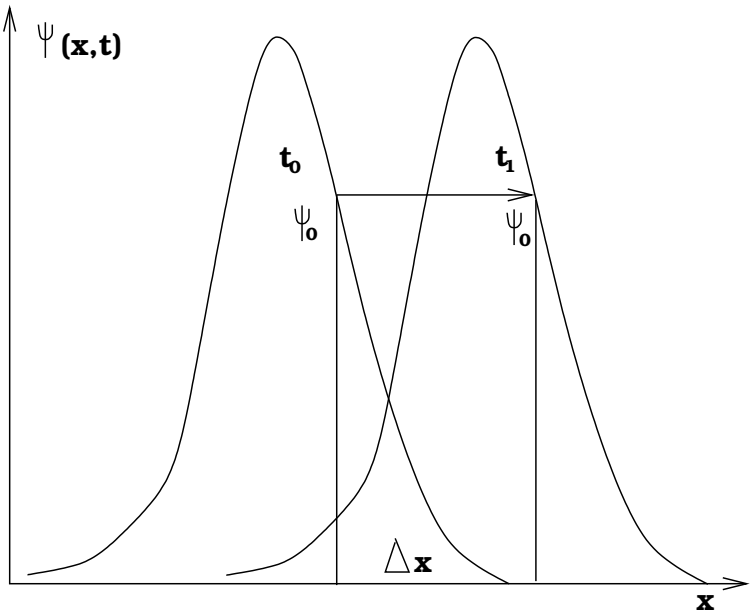}
\caption{\small The propagation of a signal $\psi(x,t)$ without a change of shape with the value $\psi_0$ of
 the amplitude as a traced attribute.}
\label{normalshape}
\end{figure}

 Let this
   attribute be labelled by some one-point fixed value $\psi_0$ of the \f $\psi(x,t)$.
 Suppose, one follows up this local "attribute" and manages a local observation of the condition:
 
 \be
 \psi(x,t)-\psi_0=0,
 \label{implicit}
 \ee

 which fixes the space-time points $\{ x, t\}$ where this 
 condition holds (Fig.\ref{normalshape}). Thus we have a \f $x(t)$ given in an implicit form (\ref{implicit}).
 
  The first order implicit derivation provides the velocity of this one-point local attribute:
 \be
 v_{(0)}(x,t):=\fr{dx}{dt}= -\fr{\fr{\pd\psi }{\pd t}}  {\fr{\pd\psi }{\pd x} },
 \label{0ord}
 \ee 
called from here on the "zero-order" or "one-point" phase velocity (0-PV).
 It describes in the simplest case the speed of translation of some arbitrary
 pulse, that can  be treated as the signal propagation velocity, provided that the measured
  value $\psi_0$ of the amplitude is the signal considered.\\
  
 To proceed with the local description of the shape propagation we now consider 
as  
  the measurable attribute of $\psi(x,t)$ at some time $t_0$ not the
single value $\psi_0=\psi(x_0;t_0)$ at the point $x_0$, but a set of values $\psi$ in a local
 neighborhood (like ${\cal U}(x)$) close to this point. 
  Then another local attribute can be constructed to trace their propagation.
 If, for instance, one looks at a certain value of the first derivative $\pd\psi(x; t_0)/ \pd x:=\chi_0$ of the
 shape $\psi$, (like $\pd\psi(x; t_0)/ \pd x=0$, i.e. at a maximum or minimum point), one should
  consider the propagation of the condition:
\be
\fr{\pd\psi(x,t)}{\pd x}=\chi_0;
\ee

this provides the propagation velocity via
\be
v_{(I)}(x,t):= -\fr{\fr{\pd^2\psi }{\pd t\pd x}}  {\fr{\pd^2\psi }{\pd x^2} },
\label{1ord}
\ee

called in view of (\ref{0ord}) the "first order" or "two-point" phase velocity (1-PV) respectively.

When tracking the propagation of a maximum (minimum), the conditions $ \fr{\pd\psi(x,t)}{\pd x}=0 $  and 
$\fr{\pd^2\psi }{\pd x^2}<0$ have to hold simultaneously. 

 This Ansatz is easily iterated to phase velocities of order $N$ interpreted as 
 the propagation velocities of higher order local shape attributes via  
 
 
 \be
 v_{(N)}(x,t):= -\fr{\fr{\pd^{N+1}\psi }{\pd t\pd x^N}}  {\fr{\pd^{N+1}\psi }{\pd x^{N+1}} },
 \label{Nord}
 \ee
 
 leading to N-th order ("N+1-point") PV, describing the propagation of higher order local shape attributes.
  
  The phase velocities so defined are obviously local features depending on space-time   
 coordinates. It has to be noted, that any given problem at hand might require a particular choice of a PV 
  allowed for by the definitions given above. The following examples elucidate these requirements and demonstrate
  the flexibility of this definitions.
  
  Before proceeding to it should be recalled, that the PV-spectrum
  has been obtained wanting to describe the propagation of an arbitrary pulse in terms of
 local attributes in a medium whose properties are dependent on several variables, especially time and space
  coordinates being the most prominent and natural examples thereof.
   Any initial shape $\psi(x_0,t_0)$ thus should be deformed during the propagation (or evolution, as typicaly 
   encountered for dispersive media). The ordinary phase velocity $v_{0}$ therefore is not a relevant
  criterion to characterize the shape propagation and one has to choose an appropriate PV $v_{(N)}$.
  
\begin{figure}[h]   
\epsfxsize=10cm\epsffile{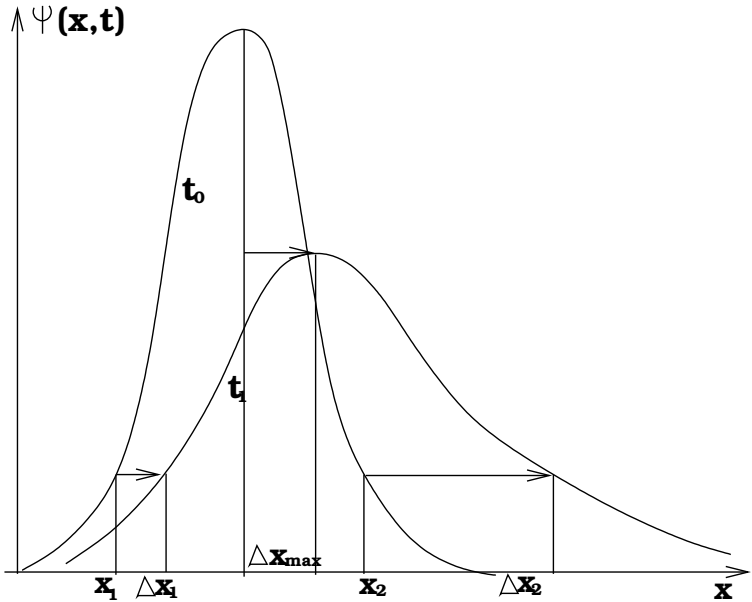}
\caption{\small A damping deformation of a signal $\psi(x,t)$. A relevant propagation attribute is the peak location (maximum)}
\label{damped}
\end{figure}

   For example, let the pulse $\psi(x,t_0)$ be subject to damping (Fig.\ref{damped}).
  As a concequence, the zeroth order PV measured in the point $x_1$ gives a magnitude much smaller
  as the same magnitude measured in the point $x_2$. 
  
\begin{figure}[h]   
\epsfxsize=10cm\epsffile{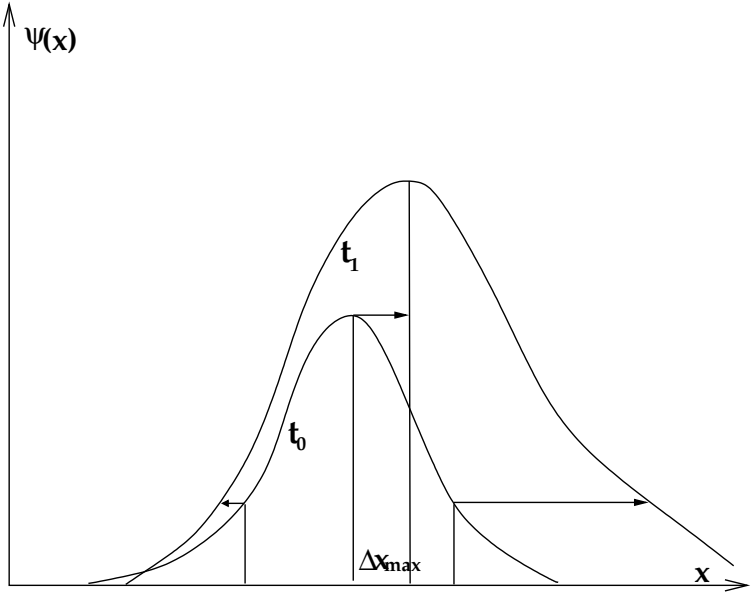}
\caption{\small The propagation of an amplified signal $\psi(x,t)$. The location of maximum describes the wave propagation}
\label{excited}
\end{figure}

   For an amplified signal (Fig.\ref{excited}, like a signal propagating in a laser excited medium), by comparison, the 
   zero order PV from the point $x_1$ provides altogether even a backward propagation.
   In both cases an appropriate approach would be to apply the first order PV     
   $v_{(I)}$ which describes the propagation of the maximum up from the point $x_0$
   properly.\\
    
\begin{figure}[h]   
\epsfxsize=10cm\epsffile{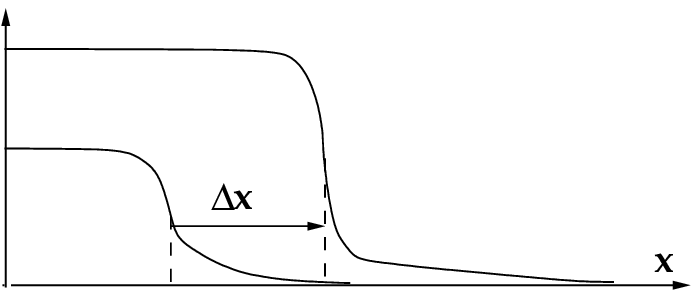}
\caption{\small The propagation of a growing kink ("tsunami model").
 The turning point is chosen to trace the propagation }
\label{kink}
\end{figure}

     For a propagation of a kink front that experiences a deformation, one 
     can check the translation of the second derivative of the shape, keeping
      track of the turning-point of the kink shape (Fig.\ref{kink}). In this case the second order ("three-point")
      PV turns out to be the relevant velocity of propagation.\\
       
 Finally, it should be noted, that the definiton of phase velocities of order zero and one,
 the  $ v_{(0)}$ and $v_{(I)}$ respectively, admits a straightforward generalization to two-,
 three-, and higher-dimensional propagation, while the phase velocities of second ( $ v_{(II)} $)
 and higher orders 
 inherently contain a certain element of ambiguity 
 in definition, since a possibility to choose
  a second-order traced attribute is not unique \cite{dimension}.   

    
\section{ Examples}
\label{ samples}

 Let us consider the conventional 1+1-dimensional wave equation 
   \be  \left[ \fr{1}{a^2}\fr{\pd^2}{\pd t^2}-\fr{\pd^2}{\pd x^2} \right]\psi(x,t)=0
   \label{free-eq}
   \ee           
   
    possessing translational solutions of the form    
    \be
    \psi(x,t)=\psi( t \pm \fr{ x}{a} ).
    \label{transl-mode} 
    \ee    
  It is easy to check that in this case the PV's of all orders defined above are
  identical and read 
  
  \be
   v_{(N)}(x,t)= a ,\ \ \ \ N=0,1,2,...
  \ee
   
  which is nothing else but the classical phase velocity  
  $\pm a$, thereby satisfying the {\it a priori} definition
   of "phase velocity" itself as a medium constant in (\ref{free-eq}).
 
 Let a propagating shape $\Psi$ now be subject to a temporal damping similar to (Fig.\ref{damped}) with
 \be
 \Psi(x,t)=\psi(t-\fr{x}{a})e^{-\la t}\equiv \psi(\phi)e^{-\la t}
 \label{t-damped}
 \ee
 
 The ordinary 0-PV velocity reads
 
 \be
 v_{(0)}=a\left( 1-\la\fr{\psi}{\psi'} \right),
 \ee
 
 where the prime denotes the derivative of $\psi$ with respect to its argument $\phi \equiv t- x/a $.
  
  This result provides a velocity with bad physical features: the velocity grows for a descending shape, for an ascending
  shape it decreases, can even be
  negative, and it diverges exactly for the peak point (under the condition that the (measured) amplitude $\Psi$ as well as
   parameter $a,\la $ have positive values). 
 
 A relevant physical velocity in this case is for instance the 1-PV 
 \be
  v_{(I)}=a\left( 1-\la\fr{\psi'}{\psi''} \right)
 \ee  
providing for a peak being traced exactly the canonical phase velocity. 
  
For this shape further PV's of higher orders are given by (\ref{Nord}): 
 \be
  v_{(N)}=a \left(1-\la \fr{\psi^{(n)}} {\psi^{(n+1)}} \right) \equiv a \left(1-\la (\log' {\psi^{(n)} })^{-1} \right) 
 \ee 
where $\psi^{(n)} $ denotes the $n$-th derivative of $\psi$ with respect to its argument as mentioned above. 
For an amplified signal as in Fig.\ref{excited} we can e.g. change the sign of $\la$.
 Especially, for the case of a kink (Fig. \ref{kink})
 
 \be
 \psi( t - \fr{x}{a} )\equiv \psi(\phi)=\arctan \phi, 
 \ee 
the spectrum of phase velocities reads by comparison :
 
 \bea
&& v_{(0)}=a\left( 1 + \la (1+\phi^2) \arctan \phi  \right) , \nn\\
&& v_{(I)}= a\left( 1 - \la \fr{1+\phi^2}{2\phi}  \right) ,\\
&& v_{(II)}= a\left( 1 - \la  \fr{\phi^3+\phi}{3\phi^2-1}  \right) .\nn 
 \eea
 
 The ordinary 0-PV has an oscillating sign at  $ \la  $ and is multiple defined because of the {\bf\it arctan} periodic \f. 
 Therefore it cannot be interpreted as a "well-defned" physical velocity.
 If the point being traced should be "labelled" by a derivative attribute,
  it turns out to be a physically inconvenient choice 
 since the shape possesses no real peaks, their propagation could be traced.
  Moreover the velocity $v_{(I)}$ diverges at $\phi=0$.
  
 The possible "labelled" attribute is also the turning-point traced by the 2-PV.
  The velocity $ v_{(II)}$  also possesses two singularities at $\phi=\pm 1/\sqrt{3}$ which do not
 coincide with the labelled point $\phi=0$, so it can be successfully followed up at the measurement.\\

 {\it Historical remark \\}
  
  Canonically the definition of velocity   
   should have proceeded with an originally artificial extension of a translational
   solution of (\ref{free-eq}):
   
   \be  \psi(x,t)=\psi( t - \fr{ x}{a} ) \equiv \psi[ \fr{1}{\om} ( \om t -\fr{\om}{a} x ) ] ,\equiv \phi( \om t -k x ),\ k\equiv \fr{\om}{a}
   \ee
  
  and the parameter $\om$ is further understood in restricted sense as a "frequency" of a "necessary" perodic harmonic \f $\phi$,
  usually $e^{\pm ix}$ as mentioned above in the introduction.  
  
   It is not surprising that the 0-order PV provides in this case the value $\om/k$, canonically interpreted as a "phase velocity of periodic wave" 
  
 In the case of any adopted interrelations between $k$ and $\om$ 
 that are not encountered in the wave equation, in particular any so called "dispersion relation"
  between frequency and wave-number,
 the PV $v_{(0)}$ is in fact a proportionality factor
  \be
d\om= v_{(0)}dk,
  \ee
 
 which is identical with the classical ("canonical") definiton of a "group velocity" \cite{brillouin}.
  
  At this point it should be recalled, that the original idea of the "group velocity" $U$, as summarized e.g. in \cite{bateman}, 
  \be \fr{\pd \la}{\pd t}  + U\fr{\pd \la}{\pd x} =0  \label{group} \ee
  was of the 
  similar form as the recent definition (\ref{0ord}) of 0-PV; the \eqq (\ref{group}) is however 
  intrinsically controversal since being constructed of local derivatives of the non-local parameter $\la$ (wavelength).
    For a variable wavelength much smaller than a vicinity of the point, this "group velocity" provides therefore a 
    natural approximation to the zero-order phase velocity $v_{(0)}$.
     This was for a long time a reason to take it for a grainted relevant physical concept.  
   
  In the present approach
  this feature appears as a physical phase velocity following in a straightforward way from the
  interpretation of phase propagation and does not require an 
  interpretation of $\om$ and $k$ as a "frequency" and "wave number", as well as a
   constancy of some "group" respectively.
   
 Note, that in the case of kink there are no suitable definitions of a "wave-group" and of a concerned "group velocity" for this solution
  since a Fourier decomposition does not work on a non-compact support.

 \section{  Lorentz covariance }
 \label{ lorentz}
 
  The ordinary zero order PV (\ref{0ord} ) possesses a important
  property, i.e. a local covariance in the sense of special relativity,
  as shown below. This is not the case for PV's
  of higher orders.
  
   Actually the ordinary PV $v_{(0)}$ measured in some stationary system $X$ takes in some other system $X'$,
    moving with a constant speed $V$, via the Lorentz transformations		  
   \be 
   x'=\fr{x-Vt}{\sqrt{1-\fr{V^2}{c^2}}},\ \  t'=\fr{t-\fr{Vx}{c^2}}{\sqrt{1-\fr{V^2}{c^2}}} 
   \ee
   
  the form

  \be
  v'_{(0)}=\fr{v_{(0)}+V}{1+\fr{v_{(0)} V}{c^2}}.
  \ee 
 
 This means that the zero order PV respects the relativistic velocity addition.
 Especially, a subluminal zero order PV remains also subluminal in any other
  moving system $X'$.
  
  This result should not be a surprise, since the definition of the $ v_{(0)}$ 
 is implied by the condition:
 \be
  \psi(x,t)=const,\ \
  \ee
  which remains to be of the same form under arbitrary transormations $x=x(x', t'); t=t(x', t')$, implying
  \be
   d\psi(x,t)\equiv \fr{\pd\psi }{\pd t}dt+\fr{\pd\psi }{\pd x}dx=0 
   \label{differential_0}
 \ee  
 
 for the first order differential form (or simply first differential), which possesses a form-invariance property under 
 transformations.
 
 Since \be v_{(0)}(x,t)=\fr{dx}{dt}\ee per definition, the \eqq (\ref{differential_0}) inplies the definition (\ref{0ord}) of 0-PV,
  so it should behave under space-time transformations as a usual velocity of a matter point.
   
 The first order PV (\ref{1ord}), by comparison, evaluated in some system $X$ 
  transformes in the moving system $X'$ to:
  \be
  v'_{(I)}= -\fr{\left\{\left( 1+\fr{V^2}{c^2} \right) \fr{\pd^2}{\pd x\pd t} - V\left(  \fr{1}{c^2}\fr{\pd^2}{\pd t^2}+ \fr{\pd^2}{\pd x^2} \right)\right\}\psi }
               {\left\{\fr{V^2}{c^4}\fr{\pd^2}{\pd t^2}  + \fr{\pd^2}{\pd x^2} -2  \fr{V}{c^2}\fr{\pd^2}{\pd t\pd x}\right\} \psi }.
  \ee
   
If the pulse $\psi (x,t)$ obeys the free wave \eqq  (\ref{free-eq}), the transformation becomes

\be
 v'_{(I)}=-\fr{ \left(1+\fr{V^2}{c^2} \right)v_{(I)}+2V }{\left(1+\fr{V^2}{c^2} \right)+\fr{2V v_{(I)}} {c^2} }
\ee
 
 i.e. a relation that should be called the "first order velocity addition". It differs obviously from the corresponding transformation
 of $v_{(0)}$, since the definition of the velocity $v_{(I)}$
 results from the condition

 \be
 d\left( \fr{\pd\psi (x,t)}{\pd x} \right)=0 
 \ee 
 whose form is explicitely non-invariant under space-time transformations.
 
 It can be shown that a subluminality of
 the first order PV is nevertheless still preserved by this transformation as well. Note, that for a signal which does not obey the
 free \eqq (\ref{free-eq}), this restriction is in general not guaranteed anymore.\\

\section{  A global velocity of signal transmission and "dynamic separation"  }
\label{ global}

  The discussion of local velocities was aimed towards an evaluation of global 
  features of signal propagation, namely the propagation through a finite spatial interval
   during a finite temporal interval.   
  In other words, a global velocity, practically measured, means roughly the length of  
 the  distance $\Delta x$ traveled by a traced attribute divided by the time interval $\Delta t$.\\
 
\begin{figure}[h]   
\epsfxsize=10cm\epsffile{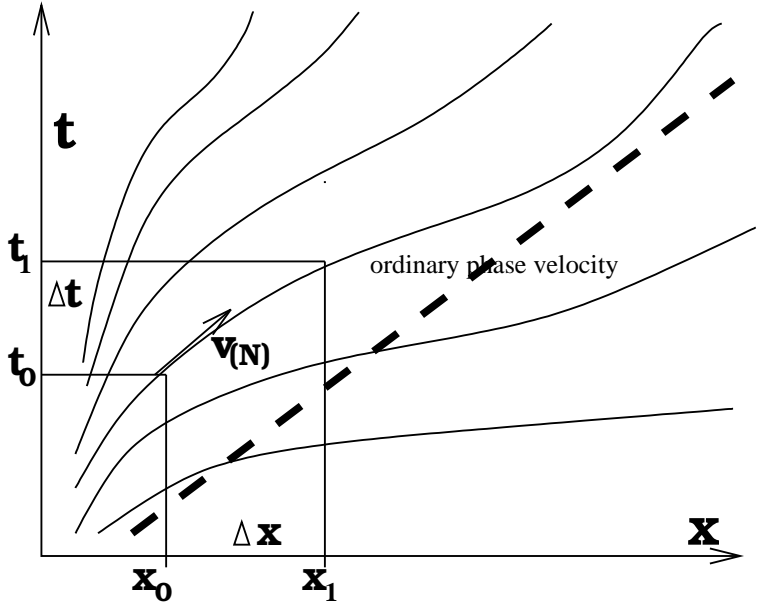}
\caption{\small Phase velocities as a family of isoclines $v_{(N)}(x,t)$ and averaged global velocities between two
events (measurements)}
\label{isoclines}
\end{figure}

 The local PV's analyzed above can be interpreted as first order differential
 \eqqs of the form 
  \be
   v_{(N)}(x,t)=\fr{dx}{dt},
   \label{def-vN}
   \ee 
  that can be illustrated graphically as a field of {isoclines} (Fig.\ref{isoclines}).

Here the local PV is the tangent function of the tangent vector of the isocline, and the averaged (total) velocity
between $(t_0,x_0)$ and  $(t_1,x_1)$ is represented by the tangent of the 
hypotenuse of the triangle $\{ (t_0,x_0),(t_1,x_1), (t_0,x_1)\}$, see Fig.\ref{isoclines}.
   
  This procedure is elucidated best 
  by some clear and well known examples:  
   Consider the propagation of a translation mode of the form

   \be
     \psi(x,t)=\psi(\xi t - k(x) x) 
   \ee  
   
   describing, for instance, an electromagnetic wave propagation in an inhomogeneous dielectric medium.
    The $k(x)$ depends now on the space coordinate, resulting from the 
 optical inhomogeneity:   
   \be
   k(x)= \xi\fr{n(x)}{c},
   \ee
where $n(x)$  is the spatially dependent index of refraction.  
    Then the local zero order PV is provided by Eq. (\ref{0ord}) and results in the first order \eqq
    
    \be
    v_{(0)}(x,t)\equiv\fr{dx}{dt}= \fr{c}{n'x+n},\ \ n'\equiv\fr{\pd n(x)}{\pd x} 
    \label{0PV_local}
    \ee     
An explicit integration of the (\ref{0PV_local}) leads to   
   
   \be
    c(t_1-t_0)= x_1n(x_1)-x_0n(x_0)
   \ee 
   
   for the interval of two events, where the signal attribute choosen for tracing
    has been checked at the time $t_0$ in the point $x_0$ and afterwards in $x_1$ at
   $t_1$.
     Assumed, the $(t_0=0,x_0=0 )$ is not a critical point (i.e. of the knot type)
     of the \eqq (\ref{def-vN}). Then $t_1=\Delta t,\ x_1=\Delta x$ and the global transition velocity
     between two points $x_0,  x_1$ is
   
     \be
     v_{(0)}(\Delta x)=\fr{c}{n(\Delta x)}
     \ee
     
     For the 1-PV velocity $ v_{(I)}$  of eq.(\ref{1ord}) the same procedure leads to
     \be
     \fr{c}{ v_{(I)}(x,t)}=  \fr{c}{ \xi} \log'((x n(x))')-(x n(x))'.
     \label{localDD}
     \ee
     
     This, \eqq straightforward to integrate, provides for the averaged
      velocity of two-point signal attribute  along the distance $\Delta x$:
     
      \be
     \fr{c}{ v_{(I)}}= n(\Delta x) -\fr{c}{ \xi\Delta x }\log (n'(\Delta x).
     \Delta x+n(\Delta x))
     \label{globalDD}
     \ee   
 It is worth to notice, that for the first order PV in media with
    a variable refraction index $n(x)$ an essential dependence on the parameter $\xi$ enters.
    
     The meaning of $\xi$ can be derived from the given form of solution $\psi$.
It can be e.g. interpreted as a frequency factor for a periodic mode or a damping factor for
evanescent one etc.     
    This parameter simply means an enumeration of solutions of a solution family (space of solutions).
The merit gained above is the separation of these solutions on the parameter $\xi$ through the 
different first (and higher)-order PV through the inhomogeneihty of media.     
    
  In the case of a periodic wave for instance, it can be interpreted as a "dispersion"  
    although an explicit frequency-dependence of $n(x)$  was not assumed.
    Moreover, the parameter $\xi$ has not necessarily to be interpreted as a "frequency" of some 
    time periodic oscillation, but rather as a time component of the time-spatial wave vector $\{\om, {\bf k}\}$
    for a translation mode (\ref{transl-mode}) in a general form, thus the considered phenomenon
    has another origin and is much more general as a conventional non-localized frequency dispersion
    $n(\om)$

     Thus we established for the first (and higher) order PV in optical
     inhomogeneous media the essential dependence on the t-component $\om$
      of the wave translation vector, especially frequency, even for local non dispersive
     media, which could be called "dynamic separation". The global averaged dynamic
    separation of eq. (\ref{globalDD}) survives as a corollary of the local separation of
     (\ref{localDD}). This phenomena does not occur for the ordinary zero order
     PV.

 \section{Concluding remarks }
 
  The inherent inconsistency of the classical subjects of "phase velocity" and "group velocity", 
  as well as "signal velocity" based therein has been discussed. 
   The inapplicability of these concepts for actual studies in photonics, near-field
   and nano-optics 
   has been shown to result from the essential non-locality of these terms.    
 
 An alternative approach for description of a propagation velocity has been proposed.
 It is strictly local and is based on the natural assumption of an ordinary measurement procedure.
  It does not need any {\it a priori} condition
 of periodicity, frequency, groups and packets or other canonical attributes. 
 
 The definition elaborated is applicable in a natural way for arbitrary. In a mathematical sense it describes
 a propagation of perturbation in any geometrical field. Examples could be: an acoustic wave as a pressure perturbation, 
 a gravitation wave on a fluid surface, a spin wave in a solid state, etc.
 In particular
 the formalism is very suited for the description of
 particle propagation in field theory, where particles are considered as field 
 perturbations.  
 
  This approach results in the series of measurable propagation velocities.      
 In zeroth order the propagation velocity coincides with the ordinary phase velocity and appears 
 to be ordinary Lorentz covariant; further application
 gives rise to generate "Lorentz covariance of higher orders".
 
  For propagation in inhomogeneous media, for instance for light in a medium with a space-dependent
   index of refraction phase velocities of higher orders exhibit 
   an essential dependence on time component of the wave vector solely as a result of inhomogenity, treated
    canonically as a "dispersion". This appears to be a more general phenomena,
    and does not presuppose any periodic frequency and dispersive properties of media.
    It should be called for this reason "dynamic separation".

 




\begin{thebibliography}{10}

\bibitem{brillouin}
L.~Brillouin  
{\it  Wave Propagation and Group Velocity }
  NEW YORK (U. A.): ACADEMIC PR., 1960. - XI,154 S. , 1960 . 
  
\bibitem{born}   
M.~Born, E.~Wolf {\it Principles of Optics},
Cambridge University Press, 1999

  
\bibitem{sherman}
 G.~C.~Sherman, K.~E.~Oughstun,
  {\it Phys.Rev.Lett., \underline{\bf 47} (20), 1451-1454 (1981)}
 
 \bibitem{guertler} 
 A.~G\"urtler, C.~Winnewisser, H.~Helm, P.~U.~Jepsen
   {\it J.Opt.Soc.Am.A, \underline{\bf 17} (1), 74-83 (2000)}
 
 \bibitem{bers}
 A.~Bers,
  {\it  Am.J.Phys.,  \underline{\bf 68}(5), 482-484 (2000)}
%

 \bibitem{shiren}
N.~S.~Shiren, {\it Phys.Rev.A,  \underline{\bf 128} (5), 2103-2112 (1962)} 
 
 \bibitem{kohmoto2002}
 T.~Kohmoto, S.~Furue, Y.~Fukui, K.~Nakayama, M.~Kunimoto, Y.~Fukuda,
 {\it Technical Digest. Summaries of papers presented at the Quantum Electronics and Laser Science Conference.\\
 Conference Edition (IEEE Cat. No.02CH37338) (2002) Vol.1 p.241 Vol.1 of (271+40 suppl.)pp.,} 
  
 \bibitem{kohmoto}
 T.~Kohmoto, H.~Tanaka, S.~Furue, K.~Nakayama, M.~Kunimoto, Y.~Fukuda,
 {\it Phys.Rev. A,   \underline{\bf 72}, 025802 (2005) } 
 
 \bibitem{centini}
 M.~Centini, M.~Bloemer, K.~Myneni, M.~Scalora, C.~Sibilia, M.~Bertolotti, G.~D'Aguanno,
 {\it Phys.Rev.E,   \underline{\bf 68} 016602 (2003) }
  
 \bibitem{bloch}
 S.~C.~Bloch,
 {\it Am.J.Phys., \underline{\bf 45} (6), 538-549 (1976)}
 
\bibitem{nussenzveig}
 H.~ M.~Nussenzveig,
{\it Causality and dispersion relations }
 New York Acad. Press 1972
  	
\bibitem{sonnenschein}
E.~Sonnenschein, I.~Rutkevich, D.~Censor,
{\it  J.Electromagneic Waves and Appl., \underline{\bf 14} (4), 563-565  (2000)}
  	
\bibitem{schriemer}
H.~P.~Schriemer, J.~Wheeldon, T.~Hall
{\it Proc.SPIE, Int.Soc.Opt.Eng.2005, 5971, 597104.1-597104.9 (2005)}

\bibitem{oughstun}
K.~E.~Oughstun, N.~A.~Cartwright
{\it J.Mod.Opt., \underline{\bf 52} (8), 1089-1104 (2005)}

\bibitem{gomez} 
A.~Gomez, A.~Vegas, M.~A.~Solano
{\it Microwave and Opt.Technol.Lett., \underline{\bf 44} (3), 302-303 (2005)}

\bibitem{dimension}
I.~Drozdov, A.~Stahlhofen, to be publishing.


\bibitem{pacher}
C.~Pacher, W.~Boxleitner, E.~Gornik
{\it Phys.Rew.B, \underline{\bf 71} (12), 125317-1-11 (2005) }

\bibitem{bateman}
H.~Bateman, 
{\it Partial Differential Equations of Mathematical Physics }
Cambridge University Press, 1969 (orig.1931) p.231 
\end{thebibliography}
\end{document}